\documentclass[twocolumn,preprintnumbers,superscriptaddress,endnote,nofootinbib,prd]{revtex4}
\usepackage{graphicx}

\usepackage[hypertex]{hyperref}

\newcommand{\lsim}{\lesssim}
\newcommand{\gsim}{\gtrsim}

\newcommand{\ord}[1]{\mathcal{O}{(#1)}}
\newcommand{\beq}{\begin{equation}}
\newcommand{\eeq}{\end{equation}}

\newcommand{\mP}{\bar{M}_{\rm P}}
\newcommand{\Lphi}{\Lambda_\phi}

\begin{document}

\pagestyle{plain}


\author{Hooman Davoudiasl
\footnote{email: hooman@bnl.gov}
}
\affiliation{Department of Physics, Brookhaven National Laboratory,
Upton, NY 11973, USA}

\title{\boldmath $B$-Decay Signatures of Warped Top-Condensation}

\author{Eduardo Pont\'{o}n
\footnote{email: eponton@phys.columbia.edu} 
}
\affiliation{Department of Physics, Columbia University, New York,
NY 10027, USA}


\begin{abstract}

We point out that the light radion $\phi$ in a recently proposed
Warped Top-Condensation Model (WTCM), can provide distinct signatures
in $b \to s \,\phi$, where the on-shell $\phi$ can decay with
displaced vertices.  We find that some of the parameter space of these
models is constrained by $B$-meson and astrophysical data.  Future
$B$-decay measurements can lead to the discovery of the WTCM.

\end{abstract}
\maketitle


The mechanism responsible for electroweak symmetry breaking (EWSB)
remains unknown.  In the Standard Model (SM), the vacuum expectation
value (vev) of the Higgs scalar results in EWSB. However, this minimal
picture is unstable against quadratically divergent
quantum corrections that would naturally pull the Higgs vev near a
large UV cutoff scale.  This puzzle is often referred to as the
hierarchy problem, with the UV scale taken to be the Planck
mass $\mP\simeq 2\times 10^{18}$~GeV.

In recent years, warped 5D models of the Randall-Sundrum (RS) type
\cite{Randall:1999ee} have been developed as a possible framework to
address the hierarchy.  These models are based on a slice of 5D Anti
de Sitter (AdS) spacetime of curvature $k$, truncated by 4D
boundaries, sometimes called the UV and the IR branes.  The hierarchy
is solved if the Higgs is localized near the IR brane and the distance
$L$ between the branes is stabilized such that $k L \sim 35$.  The
stabilization leads to the appearance of a weak scale scalar, the
radion $\phi$.

Even though warped models can address the hierarchy, another
possibility that can achieve the same end is to remove fundamental
scalars, and in particular the Higgs, from the theory.  In this case,
one can still have EWSB through the condensation of fermions, in
analogy with chiral symmetry breaking in QCD. Such models generically
require the appearance of strong interactions near the weak scale, as
is the case in technicolor scenarios.  However, the requisite strong
interactions can also be provided by the exchange of Kaluza-Klein (KK)
states in warped RS-type models.  Even though these models are
conjectured to be dual to 4D strongly-coupled theories
\cite{ArkaniHamed:2000ds,Rattazzi:2000hs}, based on the AdS/Conformal
Field Theory (CFT) correspondence \cite{Maldacena:1997re}, the
geometric 5D approach is interesting in its own right as a
quantitative framework.

A recently proposed model employs strong interactions, mediated by KK
gluons, between top quarks to provide effective dim-6 operators that
lead to top condensation and EWSB \cite{WTCM} (for a related but
different approach see Ref.~\cite{Burdman:2007sx}).  In this model,
the size $L$ of the extra dimension is set by the condensation and is
a prediction of the theory.  We will refer to this framework as the
Warped Top-Condensation Model (WTCM), in the following.  However, this
setup yields very heavy KK modes, of order 30~TeV, which are
well-beyond the reach of the CERN Large Hadron Collider (LHC).  In
this model, accommodating the measured top mass requires the addition
of extra TeV scale quarks, charged under SU(3) color, which can be
searched for.  However, this is the only direct LHC signature of this
model, apart from a composite Higgs doublet of mass around $500~{\rm
GeV}$.  Nonetheless, there is another feature of this setup that can
in principle yield distinct signatures, namely the appearance of a
GeV-scale radion field $\phi$ \cite{WTCM}. Theories whose main 
new physics signature is a light dilaton have also been studied in 
\cite{Goldberger:2007zk}, although in a different region of parameter 
space.

In this work, we point out that for $m_\phi\sim 1$~GeV, one can look
for signatures of the WTCM in $B$-decays.  In particular, decay modes
of the type $b \to s\, \phi$ can be allowed on-shell and result in
displaced vertex signatures, through $\phi$-decays.  We show that both
the loop-level and the tree-level FCNC effects can lead to non-trivial
constraints on the model and can also result in its potential
discovery, using $B$-meson decay data.  We will next briefly present
the essential properties of the radion relevant to our framework.

The radion $\phi$ arises as the pseudo-Goldstone scalar associated
with breaking of dilatation invariance, after the stabilization of the
extra dimension.  In 5D warped models \cite{Goldberger:1999uk}, $\phi$
couples to the trace of energy momentum tensor and its coupling is
suppressed by the scale
\beq
\Lambda_\phi=\sqrt{6 M_5^3/k} \, e^{-k L}, 
\label{Lamphi}
\eeq
where $M_5$ is the 5D fundamental scale and $\mP^2 = M_5^3/k$, up to
tiny $\ord{e^{-2k L}}$ corrections.  In the WTCM, $k \,
e^{-kL} \sim 15~{\rm TeV}$ over a wide range of parameters, 
while the radion mass is given by the approximate
relation~\cite{WTCM}
\beq
m_\phi \approx \frac{k}{\mP} (4~{\rm GeV}) 
\sim \left( \frac{35~{\rm TeV}}{\Lambda_{\phi}} \right) (4~{\rm GeV})~.
\label{mphi}
\eeq
Reliable gravity calculations and naturalness require that $k\lsim
M_5$, and hence we may expect $m_\phi \lsim 4$~GeV, on general
grounds, in our analysis.  Typical values are $\Lphi\sim 100~{\rm
TeV}$ and $m_{\phi} \sim 1~{\rm GeV}$ (with $kL \sim 30$).

We will next provide estimates of the rate for $b \to s\, \phi$ in the
context of WTCM. There are two types of possible contributions to this
process that we will consider in turn.  First, there is a loop
contribution, which is the analogue of the SM process with a light
Higgs \cite{GHR} instead of the radion.  Secondly, there is a {\it
tree-level} contribution that arises in 5D models that account for the
SM fermion flavor, by localizing the quarks and leptons along the
fifth dimension \cite{RSflavor}.  This effect is a consequence of the
non-universality of the radion coupling to fermions with different 5D
profiles and has recently been studied in Ref.~\cite{ATZ}.  Bulk
localized fermions provide well-motivated flavor models, and the WTCM
itself is based on this framework \cite{WTCM}.  Hence, the effects of
radion-coupling non-universalities are considered typical in this work.

Let us first look at the loop effect.  The properties of the radion
closely resemble those of the SM Higgs scalar $h$, and this allows us
to adapt existing calculations for the process $B \to X_{s} h$ for our
purposes.  We will follow the results of Ref.~\cite{GHR} in our
analysis.  For $B$ mesons, one can use the spectator quark
approximation so that hadronic matrix elements cancel in the ratio
${\rm Br}(B \to X_{s}\, \phi)/{\rm Br}(B \to X_{c} \, e {\bar \nu_e})
\approx {\rm Br}(b \to s \, \phi)/{\rm Br}(b \to c \, e {\bar
\nu_e})$, and therefore (we assume IR-localized top quarks)
\begin{eqnarray}
\frac{{\rm Br}(B \to X_{s}\, \phi)}{{\rm Br}(B \to X_{c} \, e {\bar \nu_e})}  = &&\! \! 
\frac{27 \sqrt{2} }{64 \pi^2}\frac{G_\phi\, m_b^2}{f(m_c/m_b)} 
\left|\frac{V_{st}^\dagger V_{tb}}{V_{cb}}\right|^2
\left (\frac{m_t}{m_b}\right )^4 \nonumber \\ 
\times &&\!\!\left (1 - \frac{m_\phi^2}{m_b^2}\right)^2~, 
\label{loop}
\end{eqnarray}
where fermion masses and the CKM matrix elements are in obvious
notation, $G_\phi \equiv (\sqrt{2} \Lphi^2)^{-1}$, and
$f(m_c/m_b)\simeq 0.5$.  Using $m_\phi \sim 1$~GeV and $\Lphi \sim
100$~TeV, we find
\beq
{\rm Br}(B \to X_{s}\, \phi) \sim 10^{-5}\; \; \;\;({\rm Loop}).
\label{loopvalue}
\eeq  

Next, we turn our attention to an additional contribution that can 
arise at tree-level, in warped models with 5D SM field content.  A
simultaneous resolution of the hierarchy and flavor puzzles of the SM
can be achieved in these models, making them quite well-motivated.
Quite generally, once the fermions are allowed to propagate in the 5D
bulk, their interactions with massive bulk modes, such as the radion,
are subject to non-universality.  That is, once the fermion mass
matrix is diagonalized, these bulk interactions, governed by the
overlap of various 5D wavefunctions, are not necessarily diagonal and
result in the appearance of FCNC's, that, in the case of the radion,
can be parametrized as \cite{ATZ}
 \beq
{\cal L}_{FV} = \frac{\phi}{\Lphi}({\bar d}^i_L d^j_R a_{ij} \sqrt{m_i m_j} + {\rm h.c.})~,
\label{LFV}
\eeq
where we have $i\neq j$.  Here, $a_{ij}$ is a function of the fermion
zero-mode profiles \cite{RSflavor}, parametrized by the quantity
$c\equiv M_\psi/k$; $M_\psi$ is the 5D fermion mass.  There are
analogous interactions involving the up-type quark sector.

In the WTCM, the light fermion masses arise from 5D 4-fermion
operators of the form\footnote{The implementation of a top seesaw in
the WTCM implies that the relevant operators involve a linear
combination of the RH top field and a new quark field with the same
quantum numbers as $t_{R}$.  However, this is not important in the
present discussion.} $(d_{ij}/\Lambda^{3}) \,
(\overline{\Psi}_{Q^{i}_{L}}\Psi_{q^{j}_{R}})
(\overline{\Psi}_{Q^{3}_{L}}\Psi_{t_{R}}) +{\rm h.c.}$ that lead at
low energies to a Yukawa interaction with a scalar bound state $H \sim
\overline{\Psi}_{Q^{3}_{L}}\Psi_{t_{R}}$.  The effective 4D Yukawa
matrices read $[N_{c}\bar{g}/(8\pi^2)] \, [d_{ij}/( \Lambda L)] \,
[M_{\rm KK}^2/\tilde{\Lambda}^2] f_{Q^{i}_{L} q^{j}_{R} Q_{L} t_{R}}
$, where $\tilde{\Lambda} = \Lambda \, e^{-kL}$ is the warped down
cutoff, $N_{c} = 3$ is the number of colors, $\bar{g} \approx 375/174$
is related to the dynamical quark mass generated by the condensation
mechanism, and $f_{Q^{i}_{L} q^{j}_{R} Q_{L} t_{R}}$ is an overlap
integral of four fermion wavefunctions, and is given in Section 3.3 of
Ref.~\cite{WTCM}.  Since $Q^{3}_{L}$ and $t_{R}$ are required to be
near the IR brane to trigger the condensation (e.g. $c_{Q^{3}} \approx
-2/3$, $c_{t_{R}} \approx -1/2$), the latter overlap integral is to a
good approximation proportional to the product of the light fermion
wavefunctions evaluated on the IR brane: $f_{Q^{i}_{L} q^{j}_{R} Q_{L}
t_{R}} \approx (1/2) f_{c_{Q^{i}}}(L) f_{c_{q^{j}_{R}}}(L)$, where the
zero-mode wavefunctions are given, for example, in Eq.~(2) of
Ref.~\cite{WTCM}.  It follows that with an anarchy assumption for the
arbitrary coefficients $d_{ij}$, the flavor structure of the WTCM is
similar to the widely studied RS models with fermions in the bulk.
For instance, the CKM matrix elements are approximately given by
$V^{CKM}_{ij} \sim f_{c_{Q^{i}}}(L)/f_{c_{Q^{j}}}(L)$ for $i < j$,
thus fixing the $SU(2)$ doublet localization parameters, $c_{Q^{i}}$
for $i = 1,2$ \cite{RSCKM}.  The localization parameters for the RH
fields can then be fixed so as to reproduce the fermion masses.  For
instance, an NDA estimate of the unknown coefficients of the 4-fermion
interactions gives $d_{ij} \sim 24\pi^3/n$, where $n$ is a measure of
the number of fields appearing in loops.  For $n = 10$, we find that
with $c_{Q^1} = 0.65$, $c_{Q^2} = 0.58$, $c_{u^1} = 0.6$, $c_{u^2} =
-0.4$, $c_{d^1} = 0.52$, $c_{d^2} = 0.45$, $c_{d^3} = 0.25$, the
fermion masses and CKM mixing angles can be accommodated under the
anarchy assumption.  In this case, we find $a_{23} \approx a_{32}
\approx 0.02$.  For smaller $n$, the required localization parameters
are very similar to the standard RS implementation of flavor (with all
light fermions closer to the UV brane).  Radion mediated FCNC's in
such a case have been recently studied in Ref.~\cite{ATZ}, where a
scan revealed typical values of $a_{ij} \sim 0.05$.  We will take this
as a representative value in the WTCM model.

Considering only the tree-level interaction (\ref{LFV}), for the
$\phi\, b s$ vertex, we find
\begin{eqnarray}
\hspace*{-5mm}
\frac{{\rm Br}(B \to X_{s}\, \phi)}{{\rm Br}(B \to X_{c} \, e {\bar \nu_e})}  = &&\! \! 
\frac{6 \pi^2 a_{bs}^2 }{\Lphi^2 G_F^2}
\frac{(m_s/m_b^3) (1 - m_\phi^2/m_b^2)^2}{|V_{cb}|^2 f(m_c/m_b)},  
\label{tree}
\end{eqnarray}
where $a_{bs}^{2} = |a_{23}|^{2} + |a_{32}|^2$ and we have ignored
terms of $\ord{m_s/m_b}$ [in particular, we neglect a term
proportional to ${\rm Re}(a_{23} a^*_{32})$].  Using as our reference
values $m_\phi \sim 1$~GeV, $\Lphi \sim 100$~TeV, and $a_{ij} \sim
0.05$, we find 
\beq
{\rm Br}(B \to X_{s}\, \phi)  \sim 10^{-2} \; \; \;\;({\rm Tree}).
\label{treevalue}
\eeq

Hence, we see that for typical values of WTCM parameters, the
contribution from tree-level non-universality is larger by a factor of
$\ord{10^3}$, compared to the reference loop-level estimate obtained
in Eq.~(\ref{loopvalue}).  The loop effect is, in principle, less
model dependent, since it does not rely on bulk fermion profiles and
the resulting off-diagonal radion couplings.  However, as noted
before, fermion localization in warped models is well-motivated and
provides a predictive and natural mechanism for explaining the SM
flavor puzzle.  Therefore, we adopt the point of view that, in the
context of WTCM, the $b\to s\, \phi$ channel can typically have
branching fractions of the order given in Eq.~(\ref{treevalue}).  Given
the relatively large size of the branching fraction for this mode, we
expects that it can provide distinct signatures that place bounds on
the model and can potentially lead to its discovery, using $B$-decay
data.

To go further in our phenomenological analysis, we must address the
eventual decay of the radion.  We will focus on $m_\phi \lsim 4$~GeV,
required for on-shell $\phi$ decays.  However, given Eq.~(\ref{mphi}),
this choice also coincides with the expected values of $m_\phi$ in the
WTCM. Over the narrow range $3.7~{\rm GeV}\lsim m_\phi \lsim 4$~GeV,
$\phi$ decays mostly into $\tau^+ \tau^-$ and $c {\bar c}$.  For
$m_\phi \lsim 3.7$~GeV, the most relevant decay modes of $\phi$ are
into $\mu^+\mu^-, {\bar s} s$, and $g g$.  Note that even though the
gluon $g$ is massless, the quantum trace anomaly of QCD allows for a
non-negligible $\phi g g$ coupling.  In the perturbative QCD regime,
which we roughly take to correspond to $m_\phi \gsim 1$~GeV, using the
results of Refs.~\cite{Giudice:2000av,Csaki:2007ns}, we obtain
\beq
\Gamma_{\mu^+\mu^-} : \Gamma_{{\bar s} s}: \Gamma_{gg} \simeq 
x_\mu ^2 m_\mu^2 : 3 \,x_s^2 m_s^2 : \left(\frac{b_4 \alpha_s}{2\pi}\right)^2 \! m_\phi^2\, ,
\label{phimodes1}
\eeq
where $m_s \simeq 104$~MeV, $b_4 = 25/3$ is the QCD $\beta$-function
for 4 flavors, and $\alpha_s/\pi \simeq 0.1$.  Here $x = (c_L +
c_R)_f$, where $f=\mu, \, s$, and $c_{L,R}$ parametrize the
localization of $f$ helicities in the extra dimension; quite typically
we expect $x \approx 1$.  We note that, in this work, brane kinetic
terms are assumed to be generated at the quantum level and are hence
ignored in radion couplings.

It then follows that, in the range $1~{\rm GeV}\lsim m_\phi \lsim
3.7$~GeV, the radion almost exclusively decays into the $gg$ final
state.  However, given the large suppression scale $\Lphi$ of $\phi$
couplings, we must first check that $\phi \to gg $ typically happens
within the detector.  Otherwise, the signal would resemble that of $b
\to s E\!\!\!/$.  To get an estimate, note that the lifetime
$\tau_\phi$ of the radion is roughly given by
\beq
\tau_\phi \sim \frac{32 \pi^3\Lphi^2}{b_4^2 \alpha_s^2 m_\phi^3} 
\sim \! 10^{-12}{\rm s} 
\left(\frac{\Lphi}{10^2~{\rm TeV}}\right)^2 \!
\left(\frac{1~{\rm GeV}}{m_\phi}\right),
\label{tauphi}
\eeq
where $\ord{1}$ factors have been ignored.  The above lifetime
$\tau_\phi$ roughly corresponds to a displaced vertex of
$\ord{0.3}$~mm.  Hence, we see that typical values of the parameters
we are considering result in $\phi$ decays well within the detector.
In addition, the displacement of the vertex is in a range that could
be detected \cite{Boutigny:1995ib,Cheng:1995im} as a distinct
signature for $\Lphi \gsim 100$~TeV.

Therefore, in the range $1~{\rm GeV}\lsim m_\phi \lsim
3.7$~GeV, the interaction in Eq.~(\ref{LFV}) results in a branching
fraction
\beq
{\rm Br}(b\to s\, \phi\to s \,g g)  \sim 10^{-2} \;\; \;\;({\rm WTCM}).
\label{sggWTCM}
\eeq
To extract a bound on the model parameters, we compare the above value
to that expected from the SM \cite{Harrison:1998yr}
\beq
{\rm Br}(b\to s \,g g)  \sim 10^{-3}\;\; \;\;({\rm SM}).
\label{sggSM}
\eeq
Ignoring a weak dependence on $m_\phi$, Eq.~(\ref{tree}) then 
suggests
\beq
|a_{bs}|/\Lphi \lsim (10^4~{\rm TeV})^{-1}\, .
\label{bound1}
\eeq
Below 1~GeV, the radion decay enters a non-perturbative regime.  Here,
the dominant $\phi$ decay modes are expected to be $\pi^{\pm,
0}\pi^{\mp,0}$ (or $K^{\pm, 0}K^{\mp,0}$, if kinematically allowed),
since we expect that these hadronic decays are controlled by the
coupling to gluons \cite{GHR}.  Given the inherent uncertainty in the
size of this effect, due to unknown chiral Lagrangian coefficients
\cite{GHR}, we will not present any numerical bounds here.  In any
event, the dominant signal for $2 m_\pi \lsim m_\phi \lsim 1$~GeV is
expected to be of the form $X_s \pi\pi$, with displaced $\pi \pi$
vertices.

In the narrow region $2 m_\mu< m_\phi < 2 m_\pi$, the dominant decay
modes of the radion are $\mu^+ \mu^-$ and $\gamma \gamma$.  Here,
unlike the gluons, we find that the final state photons mostly couple
through their bulk overlap with the radion.  To see this, note that
the QED $\beta$-function $b_{\rm QED} = 16/3$ below the $\tau$-charm
threshold \cite{Csaki:2007ns}, and due to the smallness of $\alpha
\simeq 1/137$, the quantum effect is negligible.  We then expect that
the radion partial widths will be given by
\beq
\Gamma_{\mu^+\mu^-} : \Gamma_{\gamma \gamma} \simeq 
m_\mu^2 :  m_\phi^2/(kL)^2.
\label{phimodes2}
\eeq
Since $m_\phi \approx 2 m_\mu$ over this mass interval and $k L\simeq
30$ in the WTCM, the di-muon final state dominates in
this narrow range of masses.  Experimental data \cite{PDG} give 
\beq
{\rm Br}(B \to s \,\mu^+ \mu^-) = 
(4.3 \pm 1.2)\times 10^{-6} \;\; ({\rm Data}).
\label{Exp}
\eeq   
Assuming that the SM is only perturbed by the new physics, one can
then demand that the size of the effect from WTCM not be larger than
the error on this measurement.  This then yields
\beq
|a_{bs}|/\Lphi \lsim (2\times 10^5~{\rm TeV})^{-1}\, ,
\label{bound2}
\eeq
which is a severe bound on the model parameters.  Note that in this
regime, if we keep $|a_{bs}| \sim 0.05$, the effective radion coupling
is pushed to values of order $10^4$~TeV. For such a high value of the
radion coupling scale, we expect $\tau_\phi \sim 16 \pi
\Lphi^2/(m_\mu^2 m_\phi) \sim 10^{-9}$~s, which means that the decay
starts to occur outside the detector.  However, this regime is beyond the
expected WTCM value $\Lphi \lsim 10^3$~TeV from Eq.~(\ref{mphi}).  On
the other hand, keeping the radion coupling near the $\ord{10^3}$~TeV
range suggests that the flavor parameters may need to be tuned.

The dominant decay mode, over the interval $15~{\rm MeV}\lsim m_\phi
\lsim 2 m_\mu$, is the di-photon mode.  This can be deduced from
Eq.~(\ref{phimodes2}) with $m_\mu \to m_e$, with $kL \lsim 30$.  Note
that near the upper end of this regime, we expect $\Lphi \sim
10^3$~TeV and $kL\sim 30$.  Given
\beq
\tau_\phi \sim 16 \pi (k L)^2\Lphi^2/m_\phi^3 
\label{diphotontau}
\eeq 
we find that $\tau_\phi \gsim 10^{-6}$~s and this mass interval will
be characterized by a $b \to s E\!\!\!/$ signal.  We then find that
the branching fraction (\ref{tree}) for this mode will be below
$10^{-4}$ for $|a_{bs}| \sim 0.05$.  Given that the typical
experimental bounds on the branching fraction for $B \to X_s \, \nu
{\bar \nu}$ are at this level \cite{PDG}, we do not expect a severe
bound in this regime from $B$-decay data.

So far, we have only addressed possible laboratory bounds on this
scenario.  However, for a sufficiently light radion, astrophysical
bounds could become important.  In particular, for $m_\phi \lsim
30$~MeV, radions are light compared to the supernova core temperature
of $\ord{30}$~MeV and can provide an energy loss mechanism for the
core.  For this to happen, the decay length of the radion, set by
$\phi \to \gamma \gamma$, must exceed $\ord{100}$~km, the typical size
of a type II supernova core.  We find this to be the case, in the
above mass range.
 
Thus, for $m_\phi \lsim 30$~MeV, astrophysical over cooling bounds
from SN 1987A can be important.  For example, light radions can be
produced in bremsstrahlung processes of the type $N N \to N N \phi$,
where $N$ denotes a nucleon.  Given a radion coupling of order
$m_N/\Lphi$, SN~1987~A bounds require $\Lphi \gsim 10^6$~TeV, where we
have adapted similar bounds on axion emission \cite{PDG} to the case
of scalar emission \cite{GR}.  This corresponds to $m_\phi \sim
100$~keV. The onset of over cooling constraints corresponds to $m_\phi
\lsim 30$~MeV and roughly $\Lphi \gsim 10^4$~TeV. We thus conclude
that the approximate range $0.1~{\rm MeV}\lsim m_\phi \lsim 30$~MeV,
is disfavored by astrophysical data on SN 1987 A. From a theoretic
point of view, one could also argue that very small values of $k/\mP$
reintroduce the kind of hierarchy that the WTCM is meant to address
and therefore such low radion masses are not well-motivated in this
framework [see Eq.~(\ref{mphi})].

In summary, we have considered the light radion phenomenology of the
recently proposed WTCM in the context of $B$ physics data.  We found
that the tree-level vertex $b s \phi$, from radion-fermion coupling
non-universality, provides the dominant contribution to $b \to s\,
\phi$.  Over much of the natural parameter space of the model, the
radion will decay into light hadrons or muons, with a measurable
displaced vertex.  Various bounds on the model parameters, based on
theoretical expectations within the SM and current experimental data,
were obtained.  A more detailed analysis of the current and future
data, using the vertex displacement information can either enhance the
bounds we have found or lead to possible discovery of the light radion
signal.  We also briefly considered possible astrophysical bounds that
may apply for very light radions.
 
\vspace*{3mm}

\acknowledgments

\vspace*{-2mm}

We thank S. Dawson and A. Soni for discussions.  The work of H.D. is
supported by the United States Department of Energy under Grant
Contract DE-AC02-98CH10886.  E.P. is supported by DOE under contract
DE-FG02-92ER-40699.


\end{document}